\documentclass[journal]{IEEEtran}

\IEEEoverridecommandlockouts

\usepackage{amsthm}
\usepackage{amsfonts}
\usepackage{amssymb}
\usepackage{mathrsfs}
\usepackage{mathtools}
\usepackage[english]{babel}
\usepackage{float}

\usepackage[pdftex]{graphicx}
\usepackage{epstopdf}
\usepackage{amsmath}
\usepackage{xcolor}
\usepackage{url}
\usepackage{multirow}
\usepackage{graphicx}
\newcommand{\expec}[1]{\mathbb{E}\left[#1\right]}
\DeclareMathOperator*{\argmin}{arg\,min}
 
\usepackage{bm}
\usepackage{balance}
\usepackage{microtype}
\usepackage[caption=false,font=footnotesize]{subfig}
\usepackage{cite}
\usepackage{comment}
\allowdisplaybreaks

\newtheorem{theorem}{Theorem}
\makeatletter
\newcommand{\settheoremtag}[1]{
  \let\oldthetheorem\thetheorem
  \renewcommand{\thetheorem}{#1}
  \g@addto@macro\endtheorem{
    \addtocounter{theorem}{-1}
    \global\let\thetheorem\oldthetheorem}
  }
\makeatother

\usepackage[switch]{lineno}
\usepackage[named]{algo}
\usepackage[noend]{algpseudocode}
\usepackage{algorithm}

\newcommand{\bsym}{\boldsymbol}

\newtheorem{proposition}[theorem]{Proposition}

\newcommand{\blue}[1]{{\color{black}{#1}}}

\begin{document}
%
\title{Octonion Phase Retrieval}

\author{Roman Jacome, Kumar Vijay Mishra, Brian M. Sadler and Henry Arguello
\thanks{R. J., and H. A. are with Universidad Industrial de Santander, Bucaramanga, Santander 680002 Colombia, e-mail: \{roman2162474@correo., henarfu@\}uis.edu.co.}
\thanks{K. V. M. and B. M. S. are with the United States DEVCOM Army Research Laboratory, Adelphi, MD 20783 USA, e-mail: kvm@ieee.org, brian.m.sadler6.civ@mail.mil.}
\thanks{This research was sponsored by the Army Research Office/Laboratory under Grant Number W911NF-21-1-0099, and the VIE project entitled ``Dual blind deconvolution for joint radar-communications processing''. K. V. M. acknowledges support from the National Academies of Sciences, Engineering, and Medicine via the Army Research Laboratory Harry Diamond Distinguished Fellowship. This work was supported by ICETEX and MINCIENCIAS through the CTO 2022-0716 under Grant 8284}
}
\maketitle
\IEEEpeerreviewmaketitle

\begin{abstract}
Signal processing over hypercomplex numbers arises in many optical imaging applications. In particular, spectral image or color stereo data are often processed using octonion algebra. Recently, the eight-band multispectral image phase recovery has gained salience, wherein it is desired to recover the eight bands from the phaseless measurements. In this paper, we tackle this hitherto unaddressed hypercomplex variant of the popular phase retrieval (PR) problem. We propose octonion Wirtinger flow (OWF) to recover an octonion signal from its intensity-only observation. However, contrary to the complex-valued Wirtinger flow, the non-associative nature of octonion algebra and the consequent lack of octonion derivatives make the extension to OWF non-trivial. We resolve this using the pseudo-real-matrix representation of octonion to perform the derivatives in each OWF update. We demonstrate that our approach recovers the octonion signal up to a right-octonion phase factor. Numerical experiments validate OWF-based PR with high accuracy under both noiseless and noisy measurements.  
\end{abstract}

\section{Introduction}
In several engineering problems pertaining to imaging \cite{augereau2017hypercomplex}, array processing \cite{le2004singular,liu2017channel}, wireless communications \cite{tarokh1999space, tirkkonen2002square,buvarp2023quaternion}, filtering \cite{ortolani2017frequency}, and neural networks \cite{kobayashi2017uniqueness,kobayashi2017fixed}, the signals of interest are hypercomplex, that is, they are elements of some algebras over the field of real numbers \cite{ell2006hypercomplex}. Unlike vector spaces that only allow addition and scalar multiplication, algebras admit both addition and multiplication between the elements of the algebra \cite{kantor1989hypercomplex}. Some common examples of hypercomplex signals include quaternions \cite{voight2021quaternion}, coquaternions or split-quaternions \cite{erdougdu2013complex}, biquaternions \cite{knus1995biquaternion}, and octonions \cite{okubo1995introduction}. Instead of tackling each dimension independently, hypercomplex signal processing exploits the corresponding algebra to process all signal dimensions jointly. The quaternion approaches have been successfully applied to color image processing \cite{le2003quaternion} where the color channels are mutually correlated via quaternion algebra. Quaternion signal processing tools have also been extended to Fourier transform \cite{ell2014quaternion}, neural networks \cite{isokawa2003quaternion}, and adaptive filtering \cite{wang2019quaternion}. Applying the Cayley-Dickson construction \cite{kantor1989hypercomplex} to quaternions for higher dimensions yields an octonion representation \cite{baez2002octonions}. In this paper, we focus on octonion-valued signals.

A recent application of octonions is multispectral image processing \cite{lazendic2018octonion}, wherein each pixel 7-color channel image has a vector-valued representation such that each channel corresponds to different complex-variable dimensions. Octonions have also been exploited for color-stereo image analysis \cite{yamni2021novel}, where two 3-color channel images are represented in a different imaginary dimension. The mutual processing along the color channels and two stereo images has been shown to improve the analysis. 
Recently, there has been broad interest in the recovery of the phase of a multispectral image, which is represented using octonion-valued signals, from its phaseless measurements \cite{katkovnik2021admm}. 

Conventional phase retrieval (PR) is a long-standing signal processing problem, where we want to recover a complex-valued signal $\bsym x \in \mathbb{C}^{n}$  given its phaseless measurements $\bsym y\in \mathbb{C}^{m}$ as $\bsym y = \vert\bsym{A x }\vert^2 $, where the known measurement matrix $\bsym A \in \mathbb{C}^{m\times n}$ is also complex-valued. This problem arises in several areas such as diffractive imaging \cite{bacca2019super}, X-ray crystallography \cite{pinilla2018phase}, astronomy \cite{fienup1987phase}, 
and radar waveform design \cite{pinilla2021banraw}. A plethora of algorithms have been proposed for precise PR solutions and the literature is too expansive to summarize here (see, e.g., 
\cite{pinilla2023unfolding} for recent surveys, and references therein). Broadly, the PR algorithms follow two approaches: either exploit prior knowledge of the signal structure or make additional measurements of the magnitude via, for example, the Fourier transform. 

In the context of hypercomplex signals, recently quaternion PR (QPR) was proposed for vision applications in \cite{chen2022phase_2}, where the signal and the measurement matrix were quaternion- and real-valued, respectively. 
This was later extended to a quaternion-valued sensing matrix in \cite{chen2022phase}, 
a quaternion Wirtinger flow  (QWF) algorithm was proposed to solve the QPR problem. The QWF is an extension of its popular complex-valued PR algorithm in \cite{candes2015phase}. Another QPR application has been reported in multiple image encryption that employs quaternion gyrator transform \cite{shao2014double}. In this work, we focus on the hitherto unaddressed octonion PR (OPR) problem that is encountered in the reconstruction of multispectral images. 

However, unlike QWF, it is not straightforward to extend Wirtinger flow (WF) \cite{candes2015phase} to octonions because octonion algebra lacks associative property. Hence, unlike quaternions, deriving Wirtinger-like derivatives for octonion-valued variables is very challenging \cite{xu2015theory, qi2022quaternion}. We address this problem by employing a pseudo-real-matrix representation \cite{rodman2014hermitian} of the octonion variables to formulate our octonion WF (OWF). We identify trivial ambiguities in OPR and derive the recovery guarantees. Our numerical experiments with synthetic as well as eight-channel multispectral image real data show accurate OPR with the proposed OWF under noiseless and noisy scenarios.


Throughout this paper, we reserve boldface lowercase, boldface uppercase, and calligraphic letters for vectors, matrices, and index sets, respectively. The set of octonion numbers is denoted as $\mathbb{O}$. We denote the transpose, conjugate, and Hermitian by $(\cdot)^T$, $(\cdot)^*$, and $(\cdot)^H$, respectively. The identity matrix of size $N\times N$ is $\mathbf{I}_N$. $||\cdot||_p$ is the $\ell_p$ norm. We denote $|\cdot|$ as the cardinality of a set, $\mathbb{E}\left[ \cdot \right]$ is the statistical expectation function, and $\mathbb{P}$ denotes the probability. The functions $\text{max}$ and $\text{min}$ output their arguments' maximum and minimum values, respectively. The sign function is defined as $\operatorname{sign}(c) = \frac{c}{|c|}$. 

 {\section{Desiderata for Octonion Algebra}}\label{sec:model}
We begin with the theoretical desiderata. An octonion number $x$ is defined as  {$x=x_0 + \sum_{i=1}^{7}x_i {e}_i$}, where $x_i$ are real-valued coefficients and $e_i$ are the \textit{octonion units} such that  {${e}_i^2=-1$ for $i=1,\dots,7$}. The conjugate is $ {x^* =x_0  - \sum_{1}^{7}x_i{e}_i}$. The `real part' of $x$ is $x_0$. Octonions are obtained from Cayley Dickson's construction of quaternions. Octonion algebra is non-associative and non-commutative, that is, for given three octonion numbers $x,y,z\in \mathbb{O}$, we have $(x\cdot y)\cdot z\neq x \cdot (y\cdot z)$ and $x\cdot y \neq y\cdot x$. The `purely imaginary' part of the octonion is $\operatorname{Im} x = \sum_{i=1}^{7}x_i {e}_i$.  {The magnitude of an octonion number is $\vert {x} \vert = \sqrt{\sum_{i=0}^7 x_i^2}$. The norm of an octonion vector $\mathbf{x}\in\mathbb{O}^{n}$ is $\Vert\mathbf{x}\Vert= \sqrt{\sum_{k=1}^{n} \vert\mathbf{x}_k\vert^2}$. For a real-valued vector $\mathbf{t} \in \mathbb{R}^n$, its $\ell_2$ norm is  $\Vert\mathbf{t}\Vert_2=\sqrt{\sum_{k=1}^n\mathbf{t}_k^2}$.} The octonion-valued Gaussian distribution is represented by $\mathcal{N}_\mathcal{O}$  {which is defined as $\mathcal{N}_{\mathcal{O}} = \mathcal{N}(0,1) + \sum_{i=1}^7\mathcal{N}(0,1)e_i$, where $\mathcal{N}(0,1)$ is standard normal distribution. The octonion Gaussian distribution of a $n$-dimensional octonion random variable is $\mathcal{N}_\mathcal{O}^{n}$, where each vector element is drawn from $\mathcal{N}_\mathcal{O}$. 
For further details on octonion algebra, we refer the interested reader to \cite{okubo1995introduction}.}

It follows from the non-associative octonion algebra that, unlike in quaternion algebra, a real-matrix representation of an octonion number does not exist. However, \cite{rodman2014hermitian} proposed a pseudo-real matrix representation that has been successfully employed by many octonion-valued signal applications  \cite{lazendic2018octonion}. To obtain this representation, define the real representation of the octonion number $x\in \mathbb{O}$ as $\aleph(x) = [x_0,x_1,\dots,x_7]^T\in \mathbb{R}^{8}$. Then, the injective mapping $\gimel : \mathbb{O} \rightarrow \mathbb{R}^{8\times 8}$ is the  {real matrix representation} of an octonion number \cite{rodman2014hermitian}: 
\par\noindent\small
\begin{equation}
    \gimel(x) =\begin{bsmallmatrix}
x_0 & -x_1 & -x_2 & -x_3 & -x_4 & -x_5 & -x_6 & -x_7 \\
x_1 & x_0 & x_3 & -x_2 & x_5 & -x_4 & -x_7 & x_6 \\
x_2 & -x_3 & x_0 & x_1 & x_6 & x_7 & -x_4 & -x_5 \\
x_3 & x_2 & -x_1 & x_0 & x_7 & -x_6 & x_5 & -x_4 \\
x_4 & -x_5 & -x_6 & -x_7 & x_0 & x_1 & x_2 & x_3 \\
x_5 & x_4 & -x_7 & x_6 & -x_1 & x_0 & -x_3 & x_2 \\
x_6 & x_7 & x_4 & -x_5 & -x_2 & x_3 & x_0 & -x_1 \\
x_7 & -x_6 & x_5 & x_4 & -x_3 & -x_2 & x_1 & x_0
\end{bsmallmatrix}.\nonumber
\end{equation}\normalsize
Both representations $\aleph$ and $\gimel$ are also easily extended to vector/matrix octonion variables  {i.e., given $\mathbf{A}\in\mathbb{O}^{m\times n}$, we have $\aleph(\mathbf{A}) \in \mathbb{R}^{8n\times m}$ and $\gimel({\mathbf{A}})\in\mathbb{R}^{8m\times 8n}$. Consider $\mathbf{x} \in \mathbb{O}^{n}$ and $\mathbf{A} \in \mathbb{O}^{m\times n}$, it holds $\aleph(\mathbf{Ax}) = \gimel(\mathbf{A})\aleph(\mathbf{x})$ and $\Vert\mathbf{x}\Vert_2 = \Vert \aleph(\mathbf{x})\Vert_2$. This allows us to convert the octonion product into a real-valued matrix/vector multiplication that obeys the octonion product rules. We later employ this representation for gradient-based algorithms for OPR.}

\section{OPR and Trivial Ambiguity}
Consider the octonion-valued signal $\bsym{x} \in \mathbb{O}^{n}$ and its phaseless measurements  $\mathbf{y} = \vert\mathbf{A}\mathbf{x}\vert^2\in \mathbb{R}_+^m$ where $\mathbf{A}\in\mathbb{O}^{m\times n}$ is the octonion-valued sensing matrix. Our goal is to recover the octonion-valued signal $\mathbf{x}$ from its phaseless measurements $\mathbf{y}$. Traditional WF for high dimensional signals would require concatenating all signal components thereby discarding any interaction between them. It is, therefore, desired to devise OPR recovery that also obeys octonion algebra.

{As in conventional PR problems, there also exists an intrinsic trivial ambiguity in OPR as explained below.} \\ 
\textbf{Trivial Ambiguity:} 
Given a unit octonion $q$, $\vert q\vert = 1$, the signal $\mathbf{x}$ scaled by a global right octonion factor  {{i.e., $q$ is right-multiplied to all the elements of signal $\mathbf{x}$}} leads to the same  {measurements}, i.e., $\vert\mathbf{Ax}q\vert^2=\vert\mathbf{Ax}\vert^2$. However, since the octonion algebra is non-commutative, we have $\vert \mathbf{A}q\mathbf{x}\vert^2\neq\vert\mathbf{Ax}\vert^2$. Our goal is to recover $\mathbf{x}$ up to a \textit{trivial ambiguity} of only on the right octonion phase factor. 

To this end, first define $\mathbf{x}=\mathbf{y}q$. We show that $\vert \mathbf{a}_\ell^H\mathbf{x}\vert^2=\vert \mathbf{a}_\ell^H\mathbf{y}\vert^2$ for all $\ell=1,\dots,n$ holds with high probability. We have
    $\vert \mathbf{a}_\ell^H\mathbf{x}\vert^2 - \vert \mathbf{a}_\ell^H\mathbf{y}\vert^2 = \langle \mathbf{x}\mathbf{x}^H-\mathbf{y}\mathbf{y}^H,\mathbf{a}_\ell\mathbf{a}_\ell^H\rangle_\mathbb{R}$. 
Further,\par\noindent\small
\begin{equation}
    \sum_
    {\ell=1}^{m} \left(\vert\mathbf{a}_\ell^H\mathbf{x}\vert^2 - \vert\mathbf{a}_\ell^H\mathbf{y}\vert^2\right)^2 \geq \left\langle \mathbf{x}\mathbf{x}^H-\mathbf{y}\mathbf{y}^H, {\sum_{\ell=1}^{m}}\mathbf{a}_\ell\mathbf{a}_\ell^H\right\rangle_\mathbb{R}.\label{eq:obj}
\end{equation}\normalsize
 To lower bound the left-hand side term -- a quadratic stochastic process -- we, therefore, employ the small ball method \cite{koltchinskii2015bounding}. Recall the following Proposition~\ref{prop:bound}: 
 \begin{proposition}[Lower bound on quadratic stochastic process]\cite[Theorem 2.1]{koltchinskii2015bounding}\label{prop:bound}
     Assume $\beta_\ell$ where $\ell=1,\dots,m$ to be independent copies of $\beta$. Denote a family of functions that satisfy a uniform small-ball estimation by $\mathcal{F}$. For a constant $\tau>0$, we have $Q_{\mathcal{F}}(\tau)=\operatorname{inf}_{f\in \mathcal{F}} \mathbb{P}\left[\vert f \vert\geq \tau \right]>0$ and based on the expectation of Rademacher process $R_{m}(\mathcal{F})=\expec{\sup_{f\in \mathcal{F}}\left\vert\frac{1}{m}\sum_{\ell=1}^{m}\varepsilon_\ell f(\mathbf{\beta_\ell})\right\vert}$, where $\{\varepsilon_\ell\}_{\ell=1}^{n}$ are independent, symmetric, binary-valued random variables $\varepsilon_\ell\in \{-1,1\}$. Then, for  probability at least $1-e^{-2t^2}$ for constant $t>0$, \par\noindent\small
     \begin{equation}
         \inf_{f\in\mathcal{F}} \left\vert\sum_{\ell=1}^{m}\varepsilon_\ell f(\mathbf{\beta_\ell})\right\vert\leq \tau^2 \left(Q_{\mathcal{F}}(2\tau) -\frac{4}{\tau}R_{m}{\mathcal{F}}-\frac{t}{\sqrt{n}}\right) .
     \end{equation}\normalsize
 \end{proposition}

  {Due to the randomness of the sensing matrix $\mathbf{A}$ and assuming that the octonion signal follows an octonion Gaussian distribution}, we employ Proposition~\ref{prop:bound} to establish the following result about the trivial ambiguity of OPR.
 \begin{theorem}[Trivial ambiguity of right-octonion phase factor]\label{thm:trivial}
     Consider the octonion variables $\mathbf{y}$ and $\mathbf{x}=\mathbf{y}q$,  {where $\mathbf{y}\sim \mathcal{N}_{\mathcal{O}}^{n}$ with $q \in \mathbb{O}$. Define the sensing matrix $\mathbf{A}\sim \mathcal{N}_\mathcal{O}^{m\times n}$ with rows $\mathbf{a}_\ell\in \mathbb{O}^{n}$ for $\ell=1,\dots,m$}. Then, with a probability $1-e^{-\frac{1}{2}\tilde{c}^2m}$ for some positive constant $\tilde{c}$, 
   $\sum_{\ell=1}^{m} (\vert\mathbf{a}_\ell^H\mathbf{x}\vert-\vert\mathbf{a}_\ell^H\mathbf{y}\vert)\geq \tilde{c} m \Vert \mathbf{xx}^H- \mathbf{yy}^H\Vert_F^2$, 
with $\tilde{c}>0$. 
 \end{theorem}
 
    \begin{IEEEproof}
See Appendix.
\end{IEEEproof}

\section{Recovery Algorithm}
\label{sec:algs}
 {Our proposed OWF algorithm performs Wirtinger-like iterations. These are similar to a gradient descent approach but also take into account the octonion algebra. However, the non-convexity of the OPR problem implies that we also suitably initialize the algorithm. To this end, we employ \blue{a} spectral initialization approach.}\\
\textbf{OWF Algorithm:} The octonion algebra is non-associative and, hence, lacks a clear definition of derivatives for octonion-valued variables \cite{bouchard2022calculus} including chain rule, high-dimensional gradients, and gradient-based methods such as the WF.  Optimization-based methods that employ octonion representation, as in singular value decomposition (SVD) \cite{lazendic2018octonion} or deep octonion neural networks \cite{wu2020deep}, usually resort to the pseudo-real-matrix representation to perform optimization over the real-valued variable. Inspired by this approach, we propose using this representation to solve the following OPR optimization:\par\noindent\small
\begin{align}
\mathbf{x}^{\ast} = \argmin_{\widetilde{\mathbf{x}}\in \mathbb{O}^{n}} \sum_{\ell=1}^{m} \left(\vert\mathbf{a}_\ell^H \widetilde{\mathbf{x}}\vert^2-\mathbf{y}_\ell\right)^2.
\end{align}\normalsize

Employing the real matrix representation, the problem of recovery of the octonion signal becomes\par\noindent\small
\begin{align}
    \mathbf{x}^{\ast} &= \aleph^{-1}\left(\argmin_{\widetilde{\mathbf{x}}\in \mathbb{R}^{8n}} \overbrace{\sum_{\ell=1}^{m} \left(\Vert\gimel\left(\mathbf{a}_\ell^H\right)\aleph(\widetilde{\mathbf{x}})\Vert_2^2-\mathbf{y}_\ell\right)^2}^{f(\widetilde{\mathbf{x}})}\right)\label{eq:problem_real}
\end{align}\normalsize
where the $\ell_2$ norm comes from the observation that the norm of an octonion variable is the norm of its real representation. 

Then, \eqref{eq:problem_real} is solved by gradient descent steps. 
  {Here, the key difference with respect to the traditional complex-valued approach lies in the gradient computation. In the complex-valued case, wherein the measurements are $\mathbf{y}=\vert\mathbf{A}\mathbf{x}\vert^2$ with the signal $\mathbf{x} \in \mathbb{C}^{8n}$ and sensing matrix $\mathbf{A}\in \mathbb{C}^{m\times 8n}$. Here, the gradient update is $\nabla_{\mathbf{x}} \left(\vert\mathbf{a}_\ell^* {\mathbf{x}}\vert^2-\mathbf{y}_\ell\right)^2= (\Vert\mathbf{a}_\ell^*\mathbf{x}\Vert^2-y_\ell)(\mathbf{a}_\ell\mathbf{a}_\ell^*)\mathbf{x}$. However, the octonion real-matrix representation in the OWF considers the interaction among all signal components, which is desired for multispectral we imaging. The OWF gradient of the cost function is 
$\nabla_{\widetilde{\mathbf{x}}} f(\widetilde{\mathbf{x}}) = \sum_{\ell=1}^{m}(\Vert\gimel\left(\mathbf{a}_\ell^H\right)\aleph(\widetilde{\mathbf{x}})\Vert_2^2-\mathbf{y}_\ell)\gimel\left(\mathbf{a}_\ell^H\right)^T\gimel\left(\mathbf{a}_\ell^H\right)\aleph(\widetilde{\mathbf{x}})$. }
Then, the OWF update process in the $i$-th iteration, where $i\in \{1,\dots,I\}$ such that $I$ is the maximum number of iterations, becomes 
    $\widetilde{\mathbf{x}}^{(i)} = \widetilde{\mathbf{x}}^{(i-1)} - \alpha \nabla f(\widetilde{\mathbf{x}}^{(i-1)})$, 
where $\alpha$ is a suitable selected gradient step size. From inverse real representation of $\widetilde{\mathbf{x}}$, the octonion signal is $\mathbf{x}^{\ast} = \aleph^{-1} (\widetilde{\mathbf{x}}^{(I)})$.\\
\textbf{Initialization:} A key step in most nonconvex PR approaches is the initialization of the algorithm because spurious points in the cost function can lead to local minima. Here, we employ the popular spectral initialization \cite{candes2015phase_2}, wherein our goal is to obtain the initial estimate of the true signal by computing the leading eigenvector of the octonion-valued matrix $\mathbf{Y} = \frac{1}{m}\sum_{\ell=1}^{m} \mathbf{y}_\ell \mathbf{a}_\ell\mathbf{a}_\ell^H \in \mathbb{O}^{n\times n}$. This may be achieved by solving an octonionic right eigenvalue decomposition. In \cite{dray1998octonionic}, this was solved for small octonion-valued matrices ($n<4$). However, \cite{dray1998octonionic} cannot be extended to larger matrices. Therefore, we propose to adapt the power method for the right quaternion eigenvalue decomposition \cite{li2019power} to compute the leading eigenvalue of $\mathbf{Y}$. 
This method employs power iterations over the real matrix representation and computes the inverse real representation operator $\aleph^{-1}(\cdot)$ to yield the equivalent octonion leading eigenvalue. 

To measure the error between the estimated octonion signal $\mathbf{x}^\ast$ and its true value $\mathbf{x}$, define the distance $d(\mathbf{x},\mathbf{x}^\ast) = \min_{z}\Vert \mathbf{x}^{\ast}-\mathbf{x}z\Vert$ where $z \in \{z \in \mathbb{O}\vert \vert z\vert = 1\}$ is only-phase octonion factor. We represent this distance in terms of the pseudo-real-matrix representation of octonions as $d(\mathbf{x}^\ast,\mathbf{x}) =  \min_{z}\Vert\aleph(\mathbf{x}^\ast)- \gimel(\mathbf{x}))\aleph(z)\Vert$, using the property $\Vert \mathbf{x}\Vert = \Vert \aleph(\mathbf{x})\Vert$. After some simple algebra, we get $d(\mathbf{x},\mathbf{x}^{\ast}) = \Vert\aleph(\mathbf{x}^\ast)- \gimel(\mathbf{x})) g(\mathbf{x}^{\ast})\Vert$, where 
$g(\mathbf{x}^{\ast}) = \operatorname{sign}\left(\left(\gimel(\mathbf{x})^T\aleph(\mathbf{x}^{\ast})\right)\left(\gimel(\mathbf{x})^T\gimel(\mathbf{x}\right)^{-1}\right)$.

\section{Numerical Experiments}\label{sec:results}
\begin{figure}[!t]
    \centering
    \includegraphics[width=0.85\linewidth]{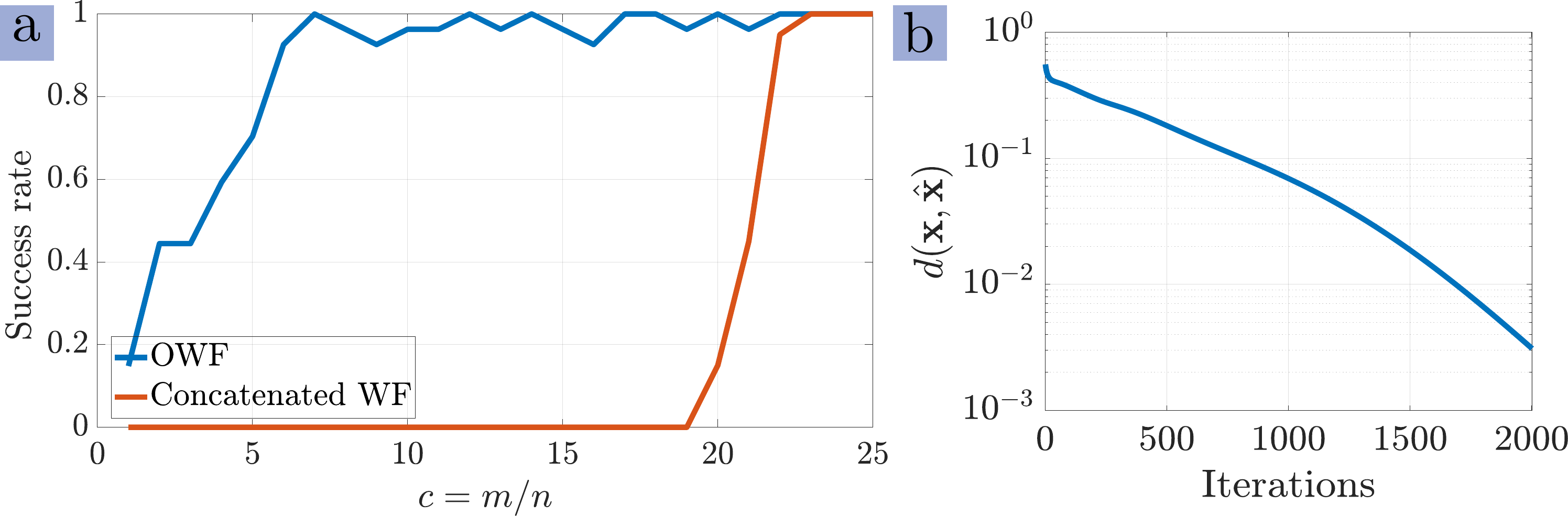}
    \caption{(a) Success rate of OWF and concatenated WF for different value of sampling complexity $m/n$ with $n=100$. (b) Convergence rate of OWF for $m/n=20$ for 2000 iterations.} 
    \label{fig:fig_curve}
\end{figure}
\begin{figure}[!t]
    \centering
    \includegraphics[width=0.70\linewidth]{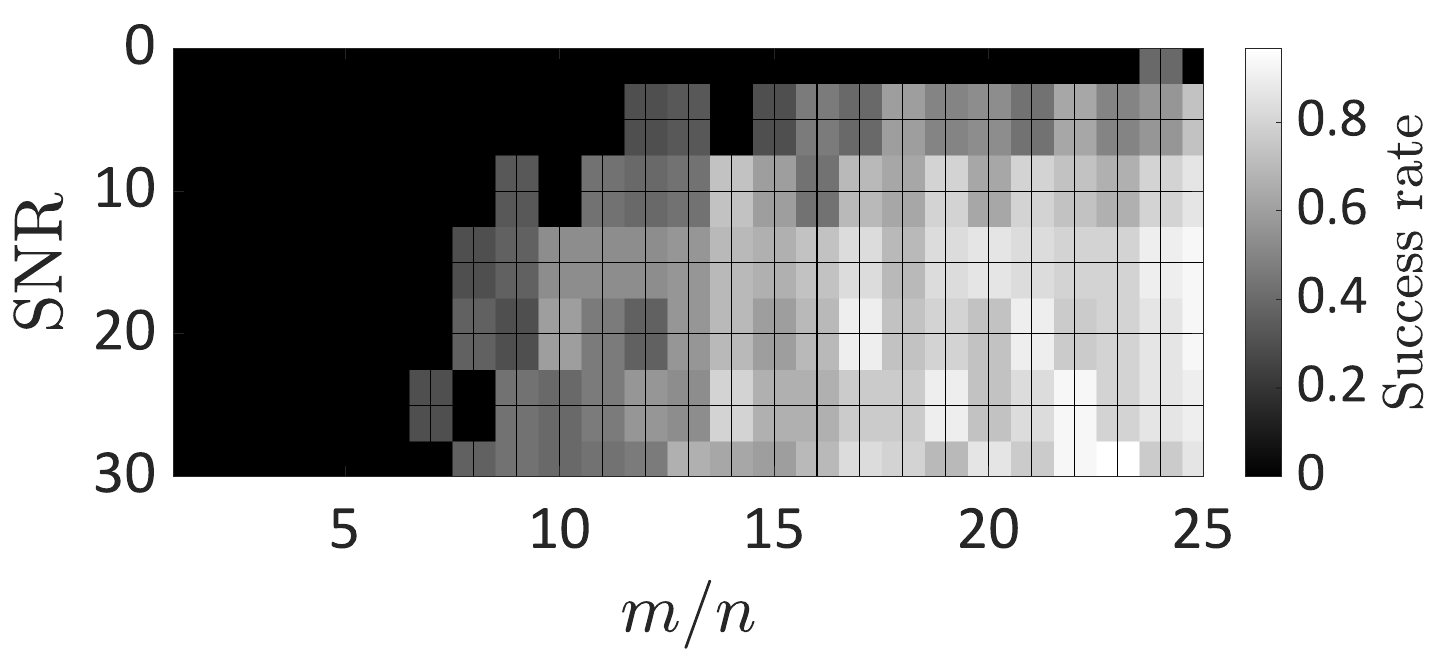} 
    \caption{Success rate of OWF for different values of sampling complexity $m/n$ with $n=30$ with measurements under additive Gaussian noise}
    \label{fig:snr}
\end{figure}
\begin{figure}[!t]
    \centering
    \includegraphics[width=0.75\linewidth]{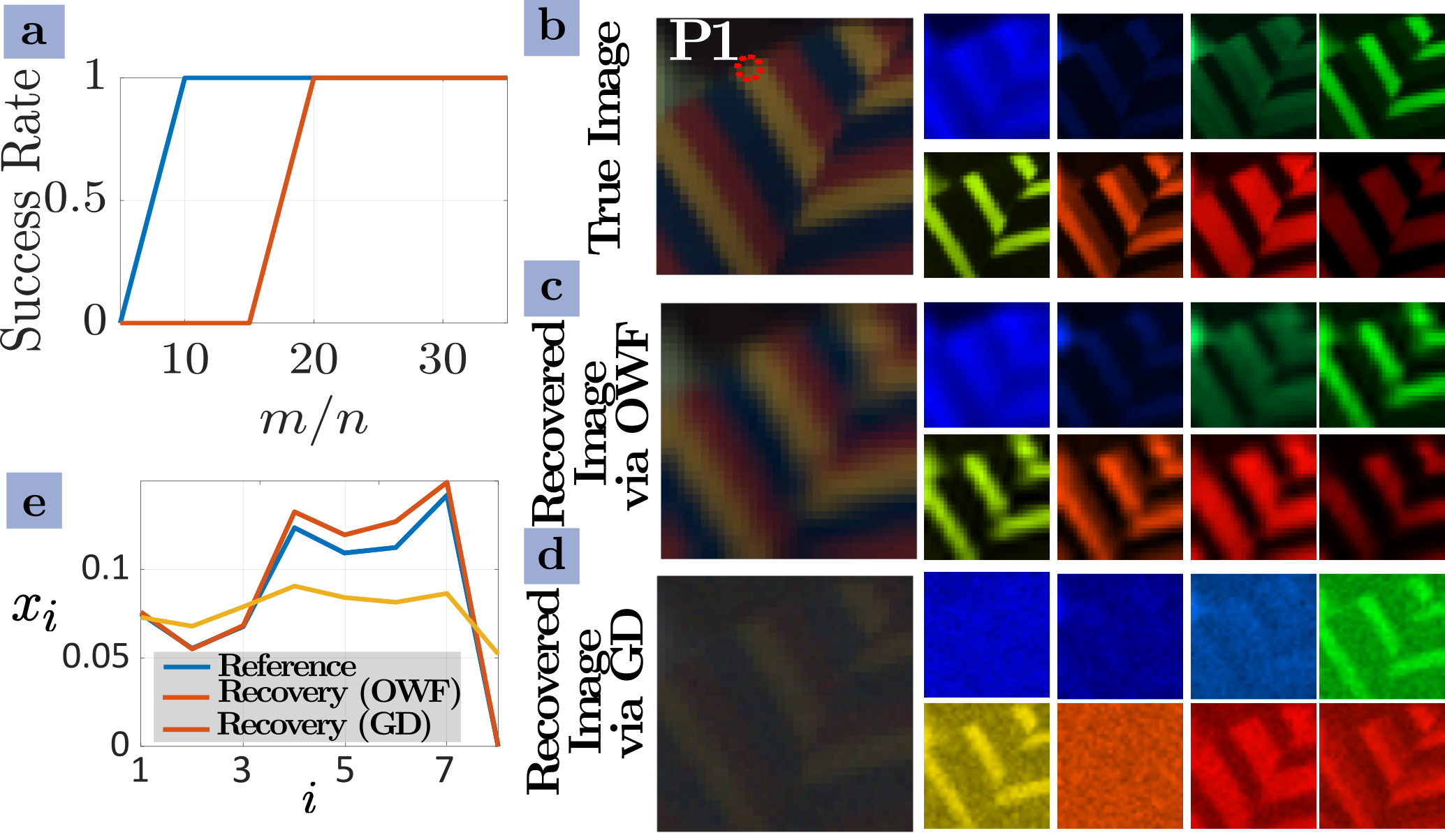}
    \caption{\blue{Reconstruction with real data. (a) Statistical performance for 64 images of the spectral image dataset. (b) RGB representation of the 8-channel spectral image and its individual 8 components on the right panel. (c) OWF-reconstructed image and its components; recovered image's PSNR $= 39.01$ dB (d) GD-reconstructed image and its components; recovered image's PSNR $= 24.16$ dB. (e) Recovered spectral signature.}} 
    \label{fig:visual}
\end{figure}

We validated our OWF algorithm 
through various numerical experiments using the quaternion and octonion toolbox for MATLAB \cite{sangwine2016quaternion}. Unless otherwise noted, the sensing matrix was drawn from an octonion Gaussian distribution i.e.,  $\mathbf{A}  {\sim} \mathcal{N}_O^{m\times n}$. The  {maximum} OWF iterations were set to $I = 2000 $. We set~$\alpha = \frac{5 m}{\sum_{\ell=1}^{m}\mathbf{y}_\ell}$ \cite{chen2022phase}. 
\\
{\textbf{Synthetic Data:}} 
{We experimented with synthetic data, wherein the signal $\mathbf{x}\in \mathbb{O}^{n\times 1}$ was generated as $\mathbf{x} \sim \mathcal{N}_\mathcal{O}^{n}$, where we normalized the signal such that $\Vert\mathbf{x}\Vert=1$ and the signal dimension $n=100$. }Over 100 Monte Carlo simulations, we declare signal recovery a ``success'' when $d(\mathbf{x},\mathbf{x}^\ast)\leq 1e^{-5}$. 
Figure \ref{fig:fig_curve} (a) shows the success rate (the mean success of the 100 experiments) for varying sample complexity $m/n$.  {We also compared OWF with the traditional WF (using PhasePack library \cite{chandra2019phasepack}), wherein the signal is concatenation of all eight components, i.e., $\mathbf{x} = [\mathbf{x}_0^T,\dots,\mathbf{x}_7^T]^T \in \mathbb{C}^{8n}$ with a random complex-valued sensing matrix  $\mathbf{A}\in \mathbb{C}^{m\times 8n}$. We used  $d(\mathbf{x},\mathbf{x}^*) = \Vert\mathbf{x}^* -\mathbf{x}\operatorname{sign}(\mathbf{x}^H\mathbf{x})\Vert$ to assess reconstruction quality. The OWF achieved almost perfect recovery for $m/n>10$ while traditional WF required $m/n>20$.}
Figure \ref{fig:fig_curve}(b) plots the distance function $d(\mathbf{x},\mathbf{x}^\ast)$ for each iteration for $m/n=20$ showing linear convergence of the OWF algorithm. 
Next, we tested the OWF algorithm for measurements corrupted by additive Gaussian noise i.e., $\mathbf{y} = \vert\mathbf{Ax}\vert^2
+\bsym\omega$ where $\bsym{\omega} \in \mathbb{R}^{m}$ is sampled from  $\bsym\omega\sim\mathcal{N}\left(0,\frac{\Vert \mathbf{y} \Vert_2^2}{10^{\frac{\textrm{SNR}}{10}}}\mathbf{I}_m\right)$. We employed the same number of iterations and $n=100$. We varied the sample complexity $m/n$  and the signal-to-noise-ratio (SNR) from 0 to 30 dB in steps of 5 dB. Figure \ref{fig:snr} demonstrates that OWF recovers the octonion signal with high accuracy with $m/n>17$ and SNR$>20$ dB. \blue{The absence of a distinct phase transition in both Figures \ref{fig:fig_curve} and \ref{fig:snr} can be attributed to the utilization of a pseudo-real matrix representation in the algorithm, as precise octonion calculus tools are unavailable.} \\ 
{\textbf{Real Data:}} We also validated OWF-based OPR with real data. We used a spectral image (Figure \ref{fig:visual}a) from the CAVE multispectral image dataset \cite{CAVE_0293}. We employ a central crop of $32\times 32$ pixel and select 8 equispaced spectral bands from the 31 original spectral bands ranging from 400 to 700 nm. Each band was vectorized and selected as a dimension of the octonion signal, thus, in this case, the octonion signal dimension was $n=1024$. \blue{We compared OWF reconstruction with the gradient descent (GD) algorithm \cite{zhang2012super}. We concatenated all color channels to form the signal $\mathbf{x}^r \in \mathbb{R}^{8n}$, the sensing matrix $\mathbf{A}^r\in \mathbb{R}^{m\times 8n}$ and, hence, the measurements  $\mathbf{y}^r=\vert\mathbf{A}^r\mathbf{x}^r\vert^2$. Note that, unlike \eqref{eq:problem_real}, this method doesn't use the real-matrix representation for the product between the rows of $\mathbf{A}^r$ ($\mathbf{a}_\ell^r$) and the signal $\mathbf{x}^r$. We employed the Lanzcos algorithm \cite{lanczos1950iteration} with $100$ power iterations for initialization. Figure \ref{fig:visual}a depicts the success rate ($d(\mathbf{x},\hat{\mathbf{x}})<10^{-3}$) for 64 images of the dataset with varying sample complexity $m/n = \{1,5,10,15,20,25,30\}$. Similar to synthetic data, the OWF (concatenation method) for real data shows perfect recovery for $m/n>10$ ($m/n>20$). 

We examined spectral image recovery with $m/n=15$ and conducted OWF for $I=2000$ iterations using the sensing matrix $\mathbf{A}$ generated as in previous experiments. Figures \ref{fig:visual}b, c, and d depict the ground-truth image, OWF reconstruction, and reconstruction via concatenation, respectively. Quality assessment utilized the peak signal-to-noise ratio (PSNR) $= 20\log\left(\frac{\operatorname{max}\left(\mathbf{x},\mathbf{x}^{\ast}\right)}{\frac{1}{8n}\sum_{i=1}^{n} \vert \mathbf{x}_i -\mathbf{x}^\ast_i\vert^2}\right)$ \cite{hore2010image}. Validation involved examining the recovery of spectral signatures, specifically vectors with eight octonion components at predefined pixel coordinates. Figure \ref{fig:visual}e demonstrates superior recovery performance with OWF over the real-valued approach with GD at coordinate (10,10) in the reference and reconstructed images.
} 

\section{Summary}\label{sec:conclusion}
We introduced an OPR algorithm for recovering 8-dimensional signals from phaseless measurements. The proposed OWF algorithm, derived from the pseudo-real-matrix representation of octonions, was validated through diverse experiments across different scenarios, sample sizes, noise levels, and real signals with multispectral images. This contributes to the advancement of hypercomplex PR applications \cite{jacome2024invitation}.

\section*{Appendix: Proof of Theorem~\ref{thm:trivial}}
 From equation \eqref{eq:obj}, we have $f(\cdot) = \vert\cdot\vert$. First, we find the small-ball estimate bound of $Q_{\mathcal{F}}(\tau)$. Define $\mathbf{W} = \mathbf{x}\mathbf{x}^H-\mathbf{y}\mathbf{y}^*$. Using Paley-Zygmund inequality \blue{\cite[Lemma 8.26]{zygmund2002trigonometric}}, we get\par\noindent\small
\begin{align}
    Q_{\mathcal{F}}(\tau) = & \inf_\mathbf{W} \mathbb{P}[\vert \langle \mathbf{W},\mathbf{a}_\ell\mathbf{a}_\ell^H\rangle_\mathbb{R}\vert^2\leq \tau]\nonumber\\ &\leq\inf_\mathbf{W} \frac{\expec{\vert \langle \mathbf{W},\mathbf{a}_\ell\mathbf{a}_\ell^H\rangle_\mathbb{R}\vert^2} -\tau^2}{\expec{\vert \langle \mathbf{W},\mathbf{a}_\ell\mathbf{a}_\ell^H\rangle_\mathbb{R}\vert^4}}, \blue{0<\tau<1}.\label{eq:qf}
\end{align}\normalsize
\blue{Then, we need to upper (lower) bound the denominator (numerator)}. Since $\mathbf{W}$ is a rank-2 matrix, we have  $\mathbf{W} = \lambda_1\mathbf{h}\mathbf{h}^H+\lambda_2\mathbf{b}\mathbf{b}^H$,
where $\lambda_1+\lambda_2=1$ and $\Vert\mathbf{h}\Vert=\Vert\mathbf{b}\Vert=1$ are normalized eigenvalues and eigenvectors, respectively. Then,$
        \langle \mathbf{W},\mathbf{a}_\ell\mathbf{a}_\ell^H\rangle_\mathbb{R} = \lambda_1\vert\mathbf{h}^H\mathbf{a}_\ell\vert^2+\lambda_2\vert\mathbf{b}^H\mathbf{a}_\ell\vert^2.$ 
Since $\mathbf{A}\sim \mathcal{N}_{\mathcal{O}}^{m\times n}$ and $\mathbf{x}\sim \mathcal{N}_{\mathcal{O}}^{n}$, then \blue{following the rotation invariance property of octonion product,} $\mathbf{h}^H\mathbf{a}_\ell$ and  $\blue{\mathbf{b}^H}\mathbf{a}_\ell$ are independent copies of the octonion Gaussian distribution. We observe that $\vert \mathbf{h}^H\mathbf{a}_\ell\vert^2$ and $\vert \mathbf{b}^*\mathbf{a}_\ell\vert^2$ conform to the $\chi^2$ distribution with 8 degrees of freedom. \blue{Leveraging computations of high-order moments \cite{nadarajah2008explicit} for the $\chi^2$ distribution and after some algebraic manipulations}, we get
    $\expec{\vert \langle \mathbf{W},\mathbf{a}_\ell\mathbf{a}_\ell^H\rangle_\mathbb{R}\vert^4} 
    \leq \frac{1}{4^4}\expec{8\vert\mathbf{h}^H\mathbf{a}_\ell\vert^2 +8\vert\mathbf{b}^*\mathbf{a}_\ell\vert^2 } 
 =\frac{2^4\Gamma(16)}{4^4 \Gamma(8)}\coloneqq c_0$.
On the other hand, 
    $\expec{\vert \langle \mathbf{W},\mathbf{a}_\ell\mathbf{a}_\ell^H\rangle_\mathbb{R}\vert^2} 
    \geq \frac{1}{16}\left(\expec{8\vert\mathbf{h}^H\mathbf{a}_\ell\vert^2} -\expec{8\vert\mathbf{b}^*\mathbf{a}_\ell\vert^2}\right)
    =\frac{1}{4}\coloneqq c_1$.
Define the constant $c = \frac{c_1}{c_0}$. \blue{Using the property ${\vert\langle\mathbf{P},\mathbf{D}\rangle\vert}\leq \sqrt{\operatorname{rank}(\mathbf{P})}\Vert\mathbf{P}\Vert_F\Vert\mathbf{D}\Vert_2$ from the real-matrix representation of octonion numbers}, we have\par\noindent\small
\begin{align}
    R_{m}&= \expec{\sup_\mathbf{W} \frac{1}{m}\sum_{\ell=1}^{m}\varepsilon_\ell \langle \mathbf{W},\mathbf{a}_\ell\mathbf{a}_\ell^H\rangle}\nonumber\leq \frac{\sqrt{2}}{n} \expec{\left\Vert \sum_{\ell=1}^{m}\varepsilon_\ell\mathbf{a}_\ell\mathbf{a}_\ell^H\right\Vert}.\label{eq:radam}
\end{align}\normalsize

We can decompose the octonion variables following the Carley-Dickson octonion construction: $\mathbf{a}_\ell=(\bsym\alpha_\ell + \bsym\beta_\ell e_2) + (\bsym\gamma_\ell + \bsym\eta_\ell e_2)e_4$. After some tedious algebra on \eqref{eq:radam}
using the aforementioned octonion representation, we obtain\par\noindent\small
\begin{align}
    \left\Vert \sum_{\ell=1}^{m}\varepsilon_\ell\mathbf{a}_\ell\mathbf{a}_\ell^H\right\Vert = &\left\Vert \sum_{\ell=1}^{m}\varepsilon_\ell((\bsym\alpha_\ell + \bsym\beta_\ell e_2) + (\bsym\gamma_\ell + \bsym\eta_\ell e_2)e_4)\cdot\right.\nonumber\\&\left.((\bsym\alpha_\ell + \bsym\beta_\ell e_2) + (\bsym\gamma_\ell + \bsym\eta_\ell e_2)e_4)^*\right\Vert.
\end{align}\normalsize
Re-arranging the terms yields\par\noindent\small
\begin{align}
    \left\Vert \sum_{\ell=1}^{m}\varepsilon_\ell\mathbf{a}_\ell\mathbf{a}_\ell^H\right\Vert =& \left\Vert \sum_{\ell=1}^{m}\varepsilon_\ell(\bsym{\alpha}_\ell\bsym{\alpha}_\ell^* + \bsym{\beta}_\ell\bsym{\beta}_\ell^*+\bsym{\gamma}_\ell\bsym{\gamma}_\ell^*+\bsym{\eta}_\ell\bsym{\eta}_\ell^*+\right. \nonumber\\&\left.2\bsym{\beta}_\ell\bsym{\alpha}^H_\ell+2\bsym{\eta}_\ell\bsym{\gamma}^H_\ell)\right\Vert.
\end{align}\normalsize
From random matrix theory, we have an upper bound on the Rademacher Gaussian series. Using \cite[Theorem 4.1.1]{tropp2015introduction} in \eqref{eq:radam} gives
$\expec{\Vert\sum_{\ell=1}^{m}\varepsilon_\ell \bsym{\alpha}_\ell\bsym\alpha_\ell^*}\leq  \sqrt{2{\sum_\ell\Vert \bsym{\alpha}_\ell\bsym{\alpha}_\ell^*\Vert} \log n}$. Upper bounding the spectral norm \cite{van2017spectral} yields $\sum_\ell\Vert \bsym{\alpha}_\ell\bsym{\alpha}_\ell^*\Vert\leq \mathcal{O}(\log\log n)$. Then, for universal constant $C$, we have $R_{m}(\mathcal{F}) \leq C_2  n\log m
$.

Putting all terms together, we have that with a probability at least $1-e^{-\frac{1}{2}c^2 m}$, we obtain \par\noindent\small
\begin{align*}
    \inf_\mathbf{W} \frac{1}{m}\sum_{\ell=1}^{m} \vert\langle \mathbf{W},\mathbf{a}_\ell\mathbf{a}_\ell^H\rangle_\mathbb{R}\vert \geq \frac{1}{16} c - C\log m- \frac{c}{32\sqrt{m}} \geq \frac{c} {64}m.
\end{align*}\normalsize
This leads to 
   $\sum_{\ell=1}^{m} (\vert\mathbf{a}_\ell^H\mathbf{x}\vert-\vert\mathbf{a}_\ell^H\mathbf{y}\vert)\geq \tilde{c} m \Vert \mathbf{xx}^*- \mathbf{yy}^*\Vert_F^2$, 
for a constant $\tilde{c}>0$. This proves that for sufficiently small $\tilde{c}$, the only trivial ambiguity is the right-octonion factor.

\bibliographystyle{IEEEtran}
\bibliography{references}

\end{document}